# Giant Gate Response of the Charge in an Electron-Lattice Condensate

Maedeh Taheri[1,2], Jordan Teeter[1,2], Topojit Debnath[3], Nicholas Sesing[4], Tina Salguero[4], Roger K. Lake[3,*], and Alexander A. Balandin[1,2,5,*]

[1]Department of Materials Science and Engineering, University of California, Los Angeles, California, 90095 USA

[2]California NanoSystems Institute, University of California, Los Angeles, California, 90095 USA

[3]Department of Electrical and Computer Engineering, University of California, Riverside, California, 92521 USA

[4]Department of Chemistry, University of Georgia, Athens, Georgia, 30602 USA

[5]Center for Quantum Science and Engineering, University of California, Los Angeles, California, 90095 USA

**Efficient electrical capacitive control is important for the next generation of ultra-low-power and ultra-fast electronics [1] and energy-storage devices [2]. Correlated electronic phases offer a powerful route to enhancing field-effect control beyond the limits of conventional capacitive gating. In such systems, modest gate voltages can couple to an order parameter, producing responses far larger than expected from the electrostatics of non-interacting carriers [3-10]. It was demonstrated that electron-electron interactions, in which the exchange and correlation energies among electrons lower the chemical potential of an electron system as the electron density increases, can significantly increase the effective capacitance over its geometric capacitance value [11, 12]. Here, we show that the electron-lattice or electron-phonon correlations in charge density wave (CDW) condensate can lead to a *giant* gate response with**

**the corresponding capacitance enhancement. This unusual phenomenon is demonstrated in the quasi-one-dimensional CDW material, orthorhombic TaS$_3$ (*o*-TaS$_3$), where the gate-induced change in CDW charge density exceeds predictions based on geometrical gate capacitance by *one to two orders of magnitude*. This "giant gating" effect arises from the coupling of the electric field to the CDW electron-lattice condensate, demonstrating a mechanism for massively amplifying gate response *via* collective electronic behavior. We quantify the effect by determining the quantum capacitance of the CDW charge and by constructing a band diagram for the gated CDW device. The obtained results can lead to an alternative strategy for continuing the downscaling of the transistor feature size in electronic technology.**

*Corresponding authors

The CDW electron-lattice condensates have long attracted interest for their rich interplay of lattice distortions and electron correlations [13-17], as well as their coexistence with other collective phases such as superconductivity [18, 19] and antiferromagnetism [20, 21]. Recent advances in the synthesis of quasi-one-dimensional (quasi-1D) and quasi-two-dimensional (quasi-2D) van der Waals materials have renewed attention to CDWs, particularly those that persist up to or above room temperature [22-26]. The physics of the CDW state exhibits formal analogies to the Bardeen, Cooper, Schrieffer (BCS) theory of superconductivity, including the formation of a condensate and a gapped ground state. Ideally, this condensate enables dissipation-less transport *via* a sliding mode; however, in real materials, impurity pinning introduces a finite threshold electric field for depinning. Once depinned, the CDW exhibits nonlinear transport, collective current flow, and phase-locked oscillations under external RF excitation, all of which serve as hallmarks of its coherence and dynamics [13-16, 27, 28].

Gate control of collective electronic phases has been a longstanding goal in condensed matter physics, motivated by the potential to surpass the fundamental limitations of conventional semiconducting devices. In ferroelectrics, for example, polarization switching can amplify the effect of gate voltage beyond geometric capacitance limits [3, 4]. For CDWs, gating was first demonstrated on the quasi-1D material NbSe$_3$ with long channel lengths from 100 μm – 800 μm, and the primary effect was to modulate the threshold voltage for CDW sliding [9]. Gate control of the CDW transition temperatures and phase switching were demonstrated with the two-dimensional (2D) materials 1T-TiSe$_2$ and 1T-TaS$_2$ [10, 22-25]. The gating capability was exploited to implement a CDW-based oscillator in 1T-TaS$_2$, operating at room temperature [23] and CDW-based memory [29, 30]. Most recently, gating of the CDW conductivity was demonstrated in the quasi-1D material Ta$_2$NiSe$_7$ [31].

Here, we focus on the direct gate modulation of the CDW condensate itself—its charge density—using back-gated, short-channel *o*-TaS$_3$ nanowires. This geometry enables us to decouple the effects of gating on normal carriers and the collective CDW state, revealing a substantial and previously unobserved amplification of charge response to gate voltage. We separate the effect of the gate voltage on the CDW charge from measurements of the gate dependence of the CDW current and the gate dependence of the Shapiro-like steps under DC and AC excitation. We find a *giant* gate response of the CDW charge that is *one to two orders of magnitude* larger than that predicted from the geometrical gate capacitance.

To probe the gate response of the CDW state, we fabricated back-gated *o*-TaS$_3$ devices using high-quality nanowires with diameters below 40 nm and short channel lengths ranging from 4 to 18 μm, comparable to the CDW coherence length in this material [32]. The use of high-purity crystals was essential for minimizing the depinning threshold field, which scales with defect density, inverse cross-sectional area, and length [15]. The resulting devices exhibited low threshold fields ($E_{th} \approx$ 0.01–0.05 kV/cm), enabling access to the sliding regime under modest bias conditions. The material for this study was synthesized via the chemical vapor transport method [33, 34] (see Methods and Extended Data Figure 1 for details on crystal structure and characterizations). The crystal of *o*-TaS$_3$ undergoes a transition to the incommensurate CDW phase at the temperature of 215 K [35-38], with an opening of a Peierls gap of 2Δ(T = 0) =120.6 meV [39] and an incommensurate CDW wavevector along the chain direction of $Q_b = 0.255b^*$, where $b^*$ is the reciprocal lattice vector [36]. Another lattice distortion was observed at *T'*~140 K, attributed to an incommensurate to commensurate transition, where the CDW wavevector becomes commensurate along the chain direction and locks to $Q_b = b^*/4$ [36].

The $o$-TaS$_3$ devices were fabricated on Si/SiO$_2$ substrates using back-gate geometry (see Methods for details). A representative device structure with multiple contacts is shown in Figure 1 (a). To confirm the presence of the CDW phase and its transport characteristics, and to verify that the Peierls transition and the CDW depinning and sliding are consistent with prior reports of bulk $o$-TaS$_3$ crystals [36, 39], we measured the current-voltage (*I-V*) characteristics of a representative $o$-TaS$_3$ device at different temperatures. As shown in Figure 1 (b), nonlinearity in the *I–V* curves emerges below the Peierls transition temperature $T_P \sim 215$ K, consistent with CDW depinning and the onset of sliding transport. The temperature of the Peierls transition was confirmed by the resistance measurements (see Extended Data Figure 2(b)). The nonlinearity weakens with increasing temperature and vanishes above $T_p$, where the device exhibits ohmic behavior. The CDW depinning threshold field was extracted from the differential resistance ($dV/dI$) as a function of source-drain bias, using $V_{th} = E_{th} \times L$, where $V_{th}$ is the threshold voltage, $E_{th}$ is the threshold field, and $L$ is the channel length. Figure 1 (c) shows the extracted $E_{th}$ values below $T_p$, including an anomaly near 150 K that likely corresponds to the secondary CDW transition previously reported in $o$-TaS$_3$ [36, 37].

To investigate the dynamics of CDW transport, we applied a combined DC and AC bias to the $o$-TaS$_3$ nanowires and analyzed the resulting current–voltage characteristics (see Extended Data Figure 3 for detailed *I-V*s and differential curves). The presence of Shapiro-like steps—a signature of coherent CDW sliding under RF excitation—was confirmed by examining the differential resistance ($dV/dI$) at $T = 180$ K (Figure 1 (d)). Peaks in $dV/dI$ appear when the RF frequency (*f* = 75 MHz) matches integer multiples of the fundamental CDW sliding frequency $f_0 = v_c/\lambda_c$, where $v_c$ is the CDW velocity and $\lambda_c$ is its wavelength along the chain. These peaks reflect phase locking between the CDW charge oscillation present in the DC current and the external AC drive. In the voltage region of the step, the two frequencies lock, and the velocity $v_c$ becomes nearly independent of the voltage. As the RF amplitude, *V$_{RF}$*, increases, additional harmonics up to the

M. Taheri, J. Teeter, T. Debnath, N. Sesing, T. Salguero, R. K. Lake*, and A. A. Balandin* - 2025

fourth harmonic become visible, indicating strong coherence in the sliding state. The CDW current and voltage at the differential resistance peaks are labeled as $I_m$ and $V_m$ where $m$ labels the harmonics of the fundamental frequency. The CDW current, $I_m$, at each harmonic was extracted by subtracting the normal carrier contribution from the total current at each voltage $V_m$ using $I_m = I(V_m) - V_m/R(0)$, where $R(0)$ is the zero-bias resistance. The collective current of a sliding CDW at a Shapiro step $m$ is related to the fundamental frequency as [13, 27]

$$I_m(V_m) = eAn_c m f_0 \lambda_c, \qquad (1)$$

where $e$ is the magnitude of the charge of an electron, $A$ is the cross-section of the sample, and $n_c$ is the charge carrier density in the CDW condensate. Figure 1 (e) shows the perfect linear scaling of the fundamental current peak $I_1(V_1)$ with $f_0$ indicating the coherence of CDW transport in our $o$-TaS$_3$ nanowires.

The CDW carrier density, $n_c$, is determined from Eq. (1) with $A = 0.017$ $\mu m^2$ and $f_0 = 75$ MHz. The CDW wavelength $\lambda_c$ varies slightly with temperature, from its commensurate value of $4b$ for temperatures ≲ 140 K to its incommensurate value of $3.92b$ at 215 K [18], so for the extraction of $n_c(T)$, we used the average value of $3.96b$. This introduces a maximum error of ±1% in the extracted values of $n_c$, which is smaller than the plotted symbol size in Figure 1 (f). The temperature dependence of $n_c$ follows the standard BCS-like behavior and is well fitted by the BCS gap interpolation formula [40]

$$n_c(T) = n_c(0) \tanh\left[\gamma \left(\frac{T_p}{T} - 1\right)^{\frac{1}{2}}\right], \qquad (2)$$

with $n_c(0) = 2.12 \times 10^{21}$ cm$^{-3}$ and $\gamma = 2.61$. Knowing the CDW carrier density and the CDW conductivity, we extracted the CDW mobility from $\sigma_c = q\mu_c n_c$ across the range of temperatures below $T_p$. The CDW conductivity $\sigma_c$ is obtained from the experimental $I_D - V_{DS}$ curves using $\sigma_c = \frac{\partial I_D}{\partial V_{DS}} \frac{L}{Wd} - \sigma_n$, where $\sigma_n$ is the conductivity of the normal carriers. The resultant CDW mobility, $\mu_c$, increases from 0.1 cm$^2$/Vs to 1.5 cm$^2$/Vs across the temperature range from 100 K to 210 K, as shown in blue trace in Figure 1 (f).

M. Taheri, J. Teeter, T. Debnath, N. Sesing, T. Salguero, R. K. Lake*, and A. A. Balandin* - 2025

We next examined how the gate voltage influences the CDW current of two representative $o$-TaS$_3$ nanowires with $L = 4$ μm and $L = 18$ μm. As shown in Figure 2 (a), the gate voltage has little effect on the normal current or the CDW threshold voltage, in contrast with earlier observations in long-channel NbSe$_3$ devices, where the primary effect of gating was to shift the threshold field [9]. In contrast, we observe a gate dependence of the CDW current at higher source-drain voltage, indicating that the gate field modulates the CDW condensate itself. Increasing the gate voltage, results in a systematic suppression of the drain current, consistent with hole-like transport in the CDW state [9, 41-43]. The relatively large gate voltages required are due to the thick gate oxide layer in the back-gated geometry. Figure 2 (b) presents the channel conductance as a function of gate voltage for one of the short-channel devices ($L = 4$ μm) at several temperatures below $T_p = 215$ K (refer to Extended Data Figure 4 for more output dependencies on different temperatures). At all temperatures, the conductance shows a distinct two-slope behavior, with a stronger gating effect appearing beyond a threshold in $V_{GS}$. This behavior was consistently observed across multiple devices. To isolate the gate influence on the collective CDW dynamics, we measured the response under simultaneous DC and RF biasing and extracted the gate dependence of the Shapiro steps. Figure 2 (c) shows the gate bias dependency of the current under AC-DC drive ($f = 75$ MHz). Shapiro-like steps are visible for the fundamental frequency (m = 1) and second harmonic (m = 2). The peaks in the differential resistance (Figure 2 (d)), are used to extract $I_1$ and $V_1$ as a function of gate bias using $I_1 = I(V_1) - V_1/R(0)$, where $R(0)$ is the zero-bias resistance value at each $V_{GS}$ (refer to the Extended Data Figure 4 (d) for more details on the slight changes of $R(0)$ vs $V_{GS}$).

To directly quantify the effect of gate voltage on the CDW condensate, we extracted the charge density from the gate-dependent current $I_1$ measured at the fundamental Shapiro step. Figure 3 (a) shows a plot of $I_1/f_0$, proportional to $n_c \lambda_c$, as a function of gate voltage. As in Figure 1 (f), the values are converted to $n_c(T)$ using Eq. (1) with $\lambda_c$ equal to its average value $3.96b$ (Figure 3 (a)). At each gate voltage, the data is fit to Eq. (2) assuming a constant $\gamma$, and the extracted zero-temperature CDW density $n_c(0)$ is plotted in Figure 3 (b). The error bars show the $\pm 1\%$ error due to the uncertainty in $\lambda_c$. The CDW density $n_c$ ($T = 0$) changes by $\sim -6.17 \times 10^{19}$ cm$^{-3}$ as the gate voltage is swept from $-75$ V to $-25$ V. The absolute magnitude of charge modulation is exceptionally large. To explore the temperature dependence of this effect, we examined $n_c$ vs. $V_{GS}$

at three temperatures in the commensurate state (Figure 3c). Since $\frac{I_1}{eAf_0} = n_c \lambda_c$, changes in $I_1$ can result from changes in the CDW density, wavelength, or a combination of both. In the commensurate phase, $\lambda_c$ is locked to the lattice, and the lattice constant is unaffected by gate voltage. Therefore, at the low temperatures ($T \lesssim 140$ K) all observed gate-induced changes in $I_1$ reflect direct modulation of $n_c$. Extracted values of $n_c$ with $\lambda_c = 4b$, are shown for $T = 110$ K, 120 K, and 140 K, in the left three panels of Figure 3 (c). Within the steep slope region of the conductance-$V_{GS}$ curves, the changes in $n_c$ with gate voltage are 2.47, 2.70, and 3.16 ($10^{19}$ cm$^{-3}$) for temperatures 110, 120, and 140 K, respectively. Multiplying these values by the $o$-TaS$_3$ channel thickness ($d \approx 37.3$ nm), they become 0.92, 1.0, and 1.2($10^{14}$ cm$^{-2}$), respectively. More gate dependency of Shapiro steps for $L = 18$ $\mu$m channel can be found in Extended Data Figure 5.

The magnitude of the gate-induced modulation of the CDW charge $n_c$ becomes more striking when expressed in terms of an effective capacitance. We define $C_{eff} = ed\frac{\delta n_c}{\delta V_{GS}}$ where $d \approx 37.3$ nm is the $o$-TaS$_3$ channel thickness. Figure 3 (d) compares $C_{eff}$ to the geometric capacitance of the back gate $C_g = \varepsilon_{ox}/t_{ox} \approx 1.12\times10^{-8}$ F/cm$^2$, revealing that the effective capacitance exceeds the gate capacitance by factors ranging from 37 to 112, depending on the temperature.

To better understand whether the observed gate-induced modulation is a surface or bulk effect, we consider the screening mechanisms that govern the penetration of the electric field into $o$-TaS$_3$. The gate field is screened by both the thermally excited normal carriers and the polarization of the CDW. We estimate the intrinsic normal carrier concentration from the standard semiconductor expression [44] $n_i = \sqrt{N_c N_v} e^{-\Delta/k_B T} = 1.9 \times 10^{16}$ cm$^{-3}$ at $T = 120$ K, assuming $m^* = m_0$ for both electrons and holes and $\Delta = 60$ meV. The reported dielectric constant parallel to the chains $\epsilon_\parallel$ is approximately $10^6 - 10^7 \epsilon_0$ in the temperature range of $100 - 180$ K [45, 46]. We also measured the dielectric response of the $L = 4$ $\mu$m channel device, in the low-frequency regime (1 KHz to 1 MHz) and found $\epsilon_\parallel \approx 10^7 \epsilon_0$, with $V_{ac} = 0.75\, V_{th}$ at $T = 120$ K (refer to Extended Data Figure 2 (e)). The value perpendicular to the chains $\epsilon_\perp$ is not known, however in other quasi-1D materials [47], the ratio $\epsilon_\parallel/\epsilon_\perp$ is reported to be approximately $10^4$. At $T = 120$ K, the Debye screening length $\lambda_D =$ $(\frac{\epsilon_\perp k_B T}{2n_i e^2})^{1/2}$ ranges from ~39 to 120 nm for values of $\epsilon_\perp$ between $100\epsilon_0$ to $1000\epsilon_0$. Since these lengths are comparable to or larger than the nanowire thickness (~ 37.3 nm), the electric field is expected to penetrate the full cross-section of the device, suggesting that the gate modulation reflects a bulk response rather than a purely surface phenomenon.

With these values, we can remove the gate oxide dependence of $C_{eff}$ by extracting the quantum capacitance of the CDW charge from $C_{Qc} = e^2 d\delta n_c/\delta\mu = ed\delta n_c/\delta\phi_s$, where $\phi_s$ is the surface potential of the $o$-TaS$_3$ at the SiO$_2$ interface. The surface potential is determined from the series combination of the oxide capacitance and the body capacitance. At flat band, the body capacitance is $C_B = \epsilon_\perp/\lambda_D$, however, since $\lambda_D > d$, we use $C_B = \epsilon_\perp/d$. The change in surface potential is related to the change in gate voltage by $\delta\phi_s = \delta V_{GS} C_{ox}/(C_{ox} + C_B)$, so that $C_{Qc} = ed\frac{\delta n_c}{\delta V_{GS}}\left(1 + \frac{C_B}{C_{ox}}\right) = C_{eff}\left(1 + \frac{\epsilon_\perp}{\epsilon_{ox}}\frac{t_{ox}}{d}\right)$. To provide a point of reference following standard convention, we write $C_{Qc}$ in terms of an effective oxide thickness via $C_{Qc} = \frac{\epsilon_{ox}}{t_{Q,eff}}$. Using the values of $C_{eff}$ from Figure 3 (d), and $\epsilon_\perp$ ranging from $100\epsilon_0$ to $1000\epsilon_0$, $t_{Q,eff}$ ranges from $0.039 - 0.39$ Å at $T = 110$ K and from $0.013 - 0.13$ Å at $T = 140$ K. Thus, the quantum capacitance of the CDW charge is huge. However, in contrast to the usual electrostatic circuit model of a FET[48], $C_{Qc}$ does not enter into the series capacitance governing the surface potential, since the change in $n_c$ results from the incorporation of electrons from the inert filled valence band into the condensate. The condensate can polarize, but it is tied to the lattice and cannot form an accumulation layer. Thus, the condensate enters into the electrostatics through its large dielectric constant, and the rest of the screening is from the low-density thermally excited normal carriers.

To comply with "Kroemer's Lemma of Proven Ignorance,"[49] we construct a band diagram for the gated CDW shown in Figure 4. Both thermally excited carriers and the CDW condensate exhibit p-type conduction, motivating a hole-like band structure with a gap at the Fermi level induced by electron-lattice coupling. In the bulk ($x \to \infty$), the Fermi level is fixed by the ground contact, maintaining equilibrium. Near the oxide interface, a finite surface potential $\phi_s$ produces an $x$-directed electric field, leading to band bending ($-e\phi(x)$). Charge neutrality tightly pins the CDW chemical potential, thus the CDW quasi-Fermi level $F_c$ follows the electrostatic potential, and the CDW gap remains symmetrical around $F_c$, since the CDW results from the coupling of electrons and holes. The field drives normal hole accumulation and electron depletion near the surface, requiring the quasi-Fermi level for normal carriers, $F_{n,p}$, to move towards the bottom of the CDW gap. $F_{n,p}$ must remain at $E_F = 0$ to prevent steady-state current, while $F_c$ may develop a gradient due to the pinned condensate's polarizable insulator-like response below threshold. Consequently, near the surface, $F_c$ and $F_{n,p}$ decouple, placing the CDW condensate and normal carriers in disequilibrium.

While the "giant gating" effect was demonstrated in a specific CDW material and device structure, we believe that the implications of our discovery are broader and can be used in silicon complementary metal-oxide-semiconductor (CMOS) technology or beyond-CMOS technologies. The performance of the CMOS transistors was improved by replacing $SiO_2$ with a highly polarizable Hf-based oxide gate insulator. With continuous miniaturization, switching energies of the field-effect transistors are becoming comparable to the correlation energies of the electron systems [48]. This creates more opportunities for the incorporation of CDW materials into the device structures for the use of the properties of electron-lattice condensates. The CDW layers do not have to be used necessarily as the active conducting channels, but used as an auxiliary for enhancing the gating function. One should also note that CDW switching is inherently fast, and GHz speeds were demonstrated in CDW devices even at room temperature [50].

In summary, our findings reveal a fundamentally new regime of electrostatic control in condensed matter systems: direct gating of the charge content in a CDW condensate. The large absolute charge modulation, that is one to two orders of magnitude above that possible with normal carriers, underscores the nontrivial coupling between an external electric field and a collective order parameter. This demonstration of field-tuned electron-lattice condensate dynamics in low-dimensional systems adds to the growing body of evidence that correlated electron materials can provide field sensitivity rooted in their many-body ground states far above that expected from single-particle systems. Our results can lead to an alternative strategy for continuing the downscaling of the transistor feature size in Si CMOS technology or beyond-CMOS technologies.

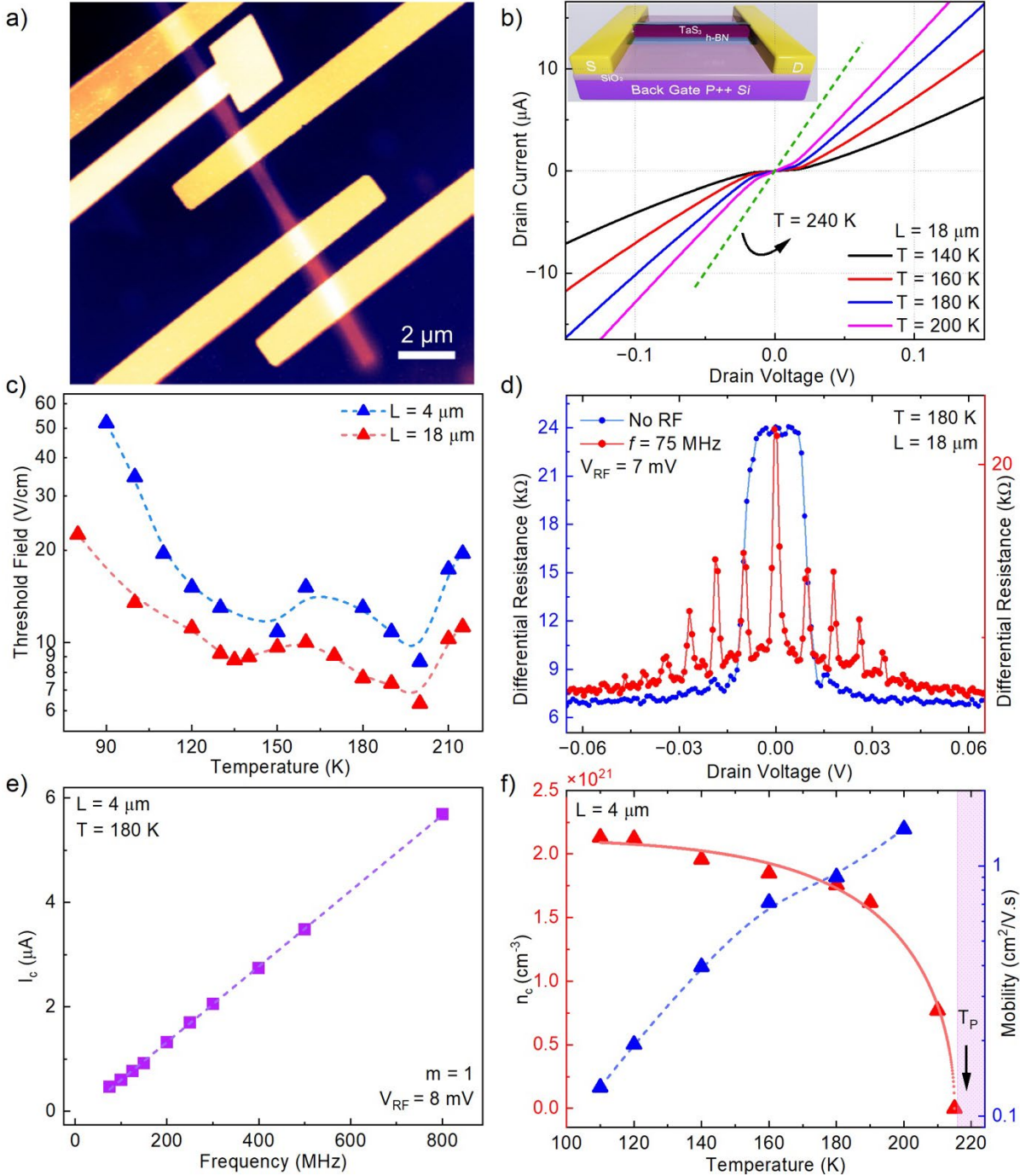

**Figure 1: Electrical characterization of the *o*-TaS$_3$ channels.** a) Atomic force microscopy (AFM) image of a quasi-1D TaS$_3$ device capped with *h*-BN b) The *I-V* characteristics of an *o*-TaS$_3$ device for temperatures ranging from $T = 140$ K to 200 K, below the transition temperature ($T_p$ ~215 K) and at $T = 240$ K above the CDW transition. c) Temperature dependence of the threshold field, $E_{th}$, below $T_p$. The anomalies around $T'_p \sim 150$ K indicate the incommensurate to commensurate transition. d) Comparison of the differential resistance $dV/dI$ with and without RF radiation. The peaks show the synchronization up to the fourth harmonic (m = 4). e) Scaling of the CDW current scales linearly with fundamental frequency at $T = 180$ K. This behavior shows the perfect coherent response for our tested TaS$_3$ devices.

M. Taheri, J. Teeter, T. Debnath, N. Sesing, T. Salguero, R. K. Lake*, and A. A. Balandin* - 2025

f) The left axis (red trace) represents CDW carrier concentration, $n_c$, as a function of temperature for $T < T_p$. Values of $n_c$ (cm$^{-3}$) for several temperatures are: $n_c$ ($T$ = 215 K) = 0, where the CDW collapses and no Shapiro step is observed, $n_c$ ($T$ = 210 K) = 7.718×10$^{20}$, $n_c$ ($T$ = 190 K) = 1.619 ×10$^{21}$, $n_c$ ($T$ = 140 K) = 1.958 ×10$^{21}$, $n_c$ ($T$ = 120 K) = 2.120 ×10$^{21}$, $n_c$ ($T$ = 110 K) = 2.132 ×10$^{21}$. The solid line represents the fitting based on BCS theory. The temperature-dependent CDW mobility is shown by the blue trace and the right axis.

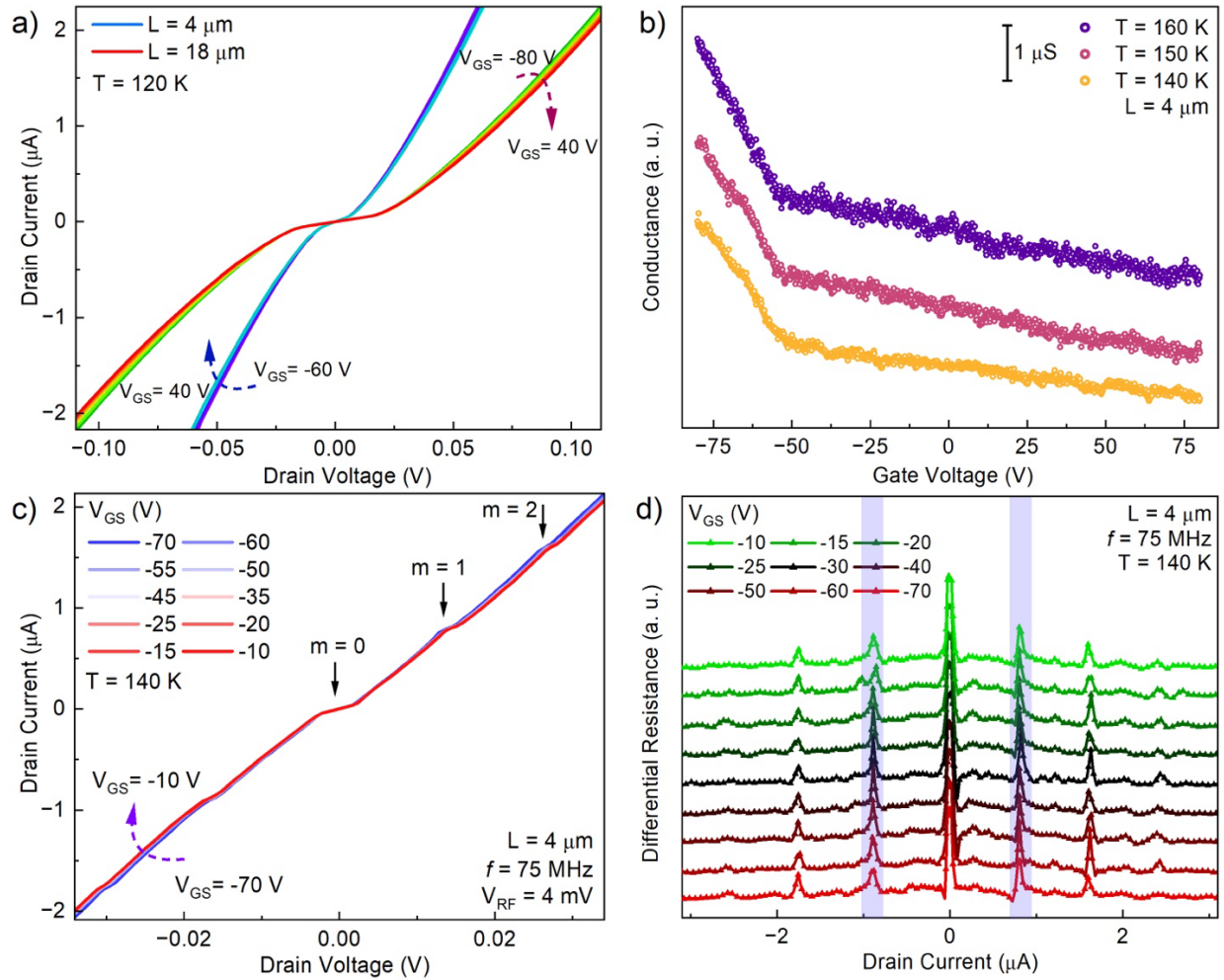

**Figure 2: The CDW-condensate field-effect devices.** a) The output $I$-$V$ characteristics for two devices, $L$ = 4 μm and $L$ = 18 μm, with cross section $A$ = 0.017 μm$^2$, for gate biases in the range of $V_{GS}$ = -80 V to 40 V in 10 V increments at $T$ = 120 K. Noticeable current modulation occurs after the threshold voltage, $V_{th}$, where the collective current shows a decrease with increasing gate bias. b) Gate modulation of the CDW conductance indicates p-type conduction with a stronger gating effect in the range of $V_{GS}$ = -80 to -50 V. c) Gate modulation of the Shapiro steps under combined AC-DC voltage. The gate dependence of the steps is determined from the differential resistance curves as a function of current across the range of gate biases, as depicted in panel (d) at $T = 140$ K.

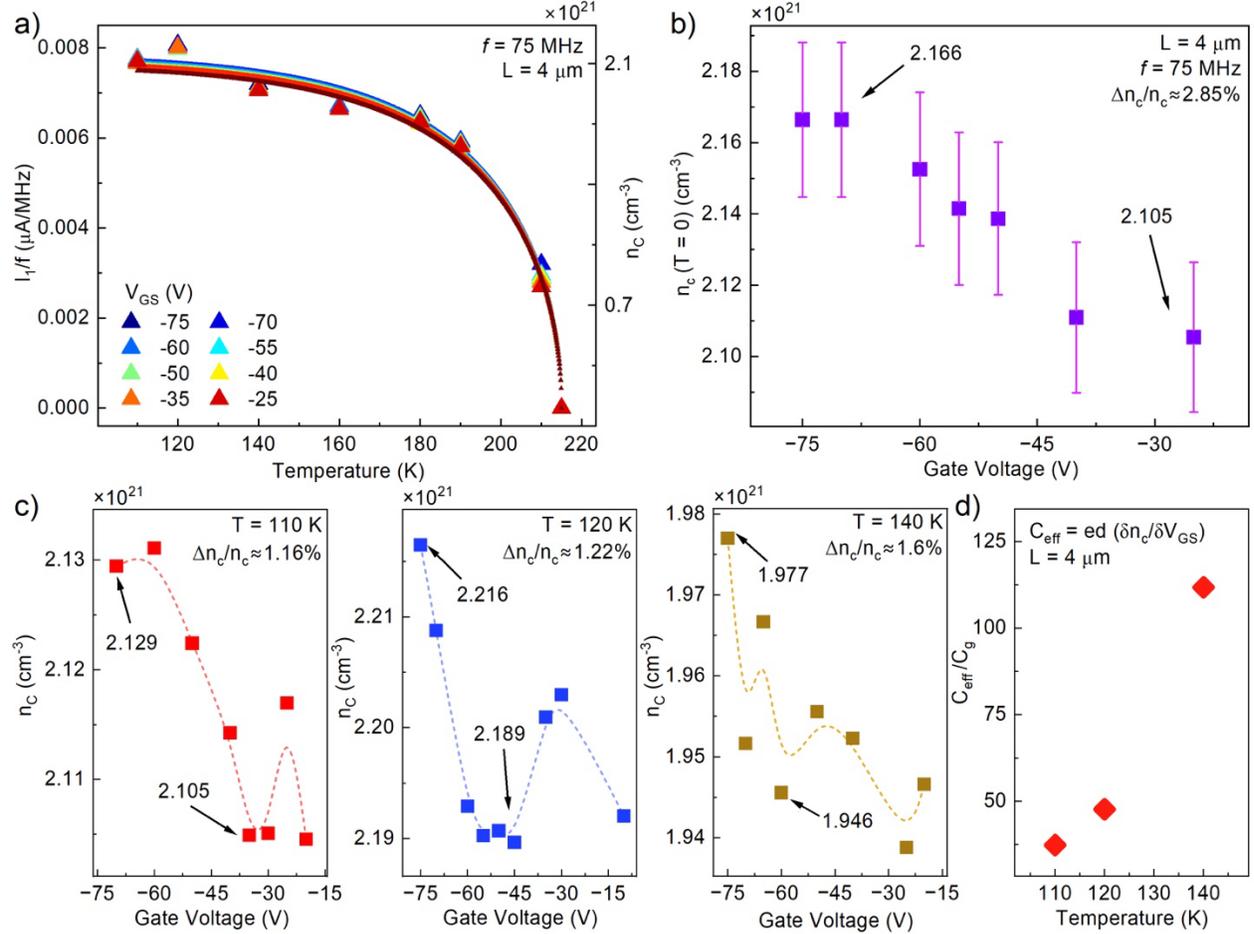

**Figure 3: Gate modulation of the CDW carrier concentration.** a) Left axis ($I_1/f$) and right axis, the extracted values of $n_c$ using the average value for $\lambda$ of $3.96\ b$. The zero-temperature value $n_c(0)$ extracted from the BCS interpolation formula (2) for various gate biases, is plotted in panel (b). The $\pm 1\%$ error bars indicate the error from the $\pm 1\%$ uncertainty in $\lambda$. The gate dependence of $n_c$ for temperatures in the commensurate phase extracted with $\lambda = 4b$ are shown in panel (c). The arrows show the minimum and maximum of the carrier density in the steep slope region of the conductance-$V_{GS}$ curves. The ratio of the effective capacitance to the geometric capacitance $C_{eff}/C_g$ plotted in panel (d) shows that $C_{eff}$ can be over two orders of magnitude larger than geometrical capacitance, $C_g$. The obtained values of $C_{eff}/C_g$ are 37, 48 and 112 for $T = 110$ K, 120 K, and 140 K, respectively.

M. Taheri, J. Teeter, T. Debnath, N. Sesing, T. Salguero, R. K. Lake*, and A. A. Balandin* - 2025

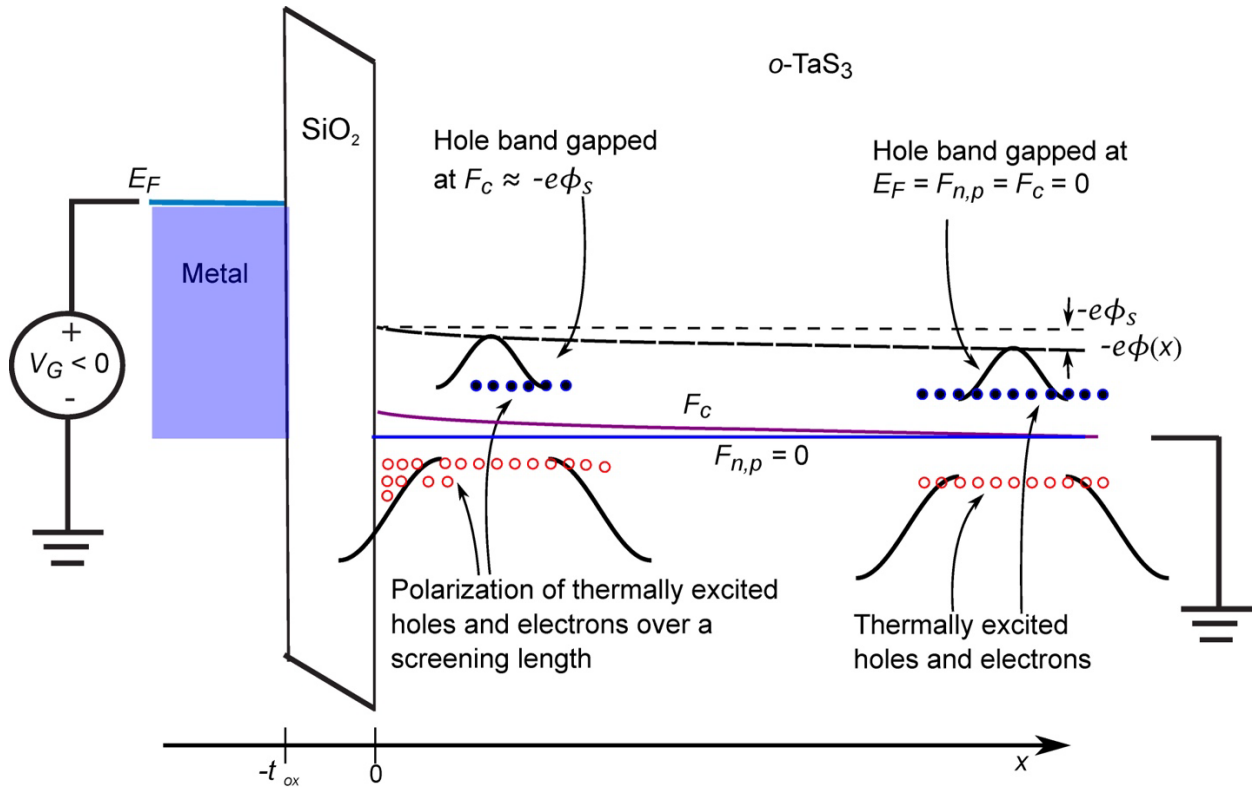

**Figure 4: Band diagram of the gated CDW channel.** The band diagram shows hole-like bands gapped by electron-lattice coupling. A surface potential induces band bending, leading to normal hole accumulation and electron depletion near the oxide interface. The CDW quasi-Fermi level, $F_c$, tracks the electrostatic potential, while the normal carrier quasi-Fermi level, $F_{n,p}$, remains flat at $E_F = 0$, resulting in disequilibrium between the condensate and normal carriers.

M. Taheri, J. Teeter, T. Debnath, N. Sesing, T. Salguero, R. K. Lake*, and A. A. Balandin* - 2025

**METHODS**

**Crystal growth and material characterization of *o*-TaS$_3$:** The high-quality *o*-TaS$_3$ crystals are synthesized using the chemical vapor transport (CVT) technique. This monoclinic polymorph with space group C2/m [1] has traditionally been referred to as 'orthorhombic' or *o*-TaS$_3$ in the old literature, and we have used this nomenclature to preserve continuity with the large body of prior work. On the benchtop, 1.9606 g (10.835 mmol) of Ta powder (Strem, 99.98%) and 1.0425 g (32.512 mmol) of S powder (J.T. Baker, ≥ 99.5%) were gently mixed. I$_2$ crystals (55.6 mg, 0.438 mmol, JT Baker, 99.9%) were placed at the bottom of a pre-cleaned and dried fused quartz ampule (~18 × 1 cm length, 10 mm inner diameter, 14 mm outer diameter, ~13 cm$^3$ volume). The mixed Ta and S powders were then transferred into the ampule, using a glass funnel and anti-static brush to ensure clean and complete transfer. After loading, the ampule was capped with an adapter and valve, then submerged in an acetonitrile/dry ice bath and subjected to four evacuation/backfilling cycles with Ar(g) using a Schlenk line. The ampule was subsequently sealed under vacuum. The sealed ampule was placed in a horizontal tube furnace, where the temperature was ramped over 10 h to establish a thermal gradient of 650 °C (source zone) to 550 °C (growth zone). This gradient was maintained for 168 h. Then the ampule was cooled to room temperature over 24 h. Upon completion, 2.8910 g of gray, densely packed, flexible wire-like crystals were recovered from throughout the ampule and left to sit in a fume hood for 1 h (96.289% isolated yield). These crystals were stored in an Ar-filled glovebox. Lattice constants were experimentally confirmed using Rietveld refinement of powder x-ray diffraction (PXRD) data. The Rietveld refined parameters (compared with literature values) are (in Å): *a* = 19.9077971 (19.9), *b* = 3.3376854 (3.34), and *c* = 15.1694670 (15.2), and the crystallographic angle *β* = 112.3662° (112.4°). Furthermore, the quality of samples has been confirmed with energy-dispersive X-ray spectroscopy (EDX) and scanning electron microscopy (SEM).

**Fabrication and electrical characterization of *h*-BN capped *o*-TaS$_3$ devices:** Thin wires with the nanometer-scale cross-section dimensions were mechanically exfoliated on the ~300 nm SiO$_2$/Si substrate. To avoid oxidation and environmental exposure, the selected nanowires were capped with thin films (thickness of ~15 to 20 nm) of hexagonal boron nitride (*h*-BN) utilizing the dry transfer method. The source-drain pattern has been fabricated with e-beam lithography (JEOL

JSM-6610), followed by the plasma etching of *h*-BN layers (Oxford 80+) and metallization of Ti/Au (15 nm / 80 nm) using e-beam evaporation (CHA Mark-40). The field-effect test structure was formed in the bottom-gate geometries, using the dry thermal $SiO_2$ as the back-gate dielectric and silver as a gate contact to the P++ silicon substrate. The devices have been tested and measured using a cryogenic system (Lakeshore TTPX) along with a semiconductor device analyzer (Agilent B1500A) with multiple source-measure units (SMU). The ohmic contact resistance of the devices was measured using four-terminal configuration with the value of ~95 Ω, which is negligible compared to channel resistance. The measurement of field-effect devices is performed using three-terminal setup, supplying voltage source, and sensing the DC current. During the measurement, the gate leakage current was constantly monitored and verified to be four-orders of magnitude smaller than the source-drain current (< 0.1 nA). For AC-DC measurement, the DC supply was coupled to the RF source, signal generator (Rigol DSG815), using a bias tee (Sigatek Bias Tee SB12D2). The low-frequency dielectric response of the channel was probed using an LCR meter (Agilent 4284A) for samples mounted in the TTPX cryogenic system.

**REFERENCES FOR METHODS SECTION**

**Acknowledgments**

The work at UCLA was supported by the Vannevar Bush Faculty Fellowship (VBFF) to A.A.B. under the Office of Naval Research (ONR) contract N00014-21-1-2947 on One-Dimensional Quantum Materials. The work at UCR and the University of Georgia was supported, in part, *via* the subcontracts of the ONR project N00014-21-1-2947. HRTEM was performed using the JEOL 2100PLUS microscope, acquired with funding from the National Institutes of Health through grant 1S10OD034282-01. The nanofabrication of the test structures was performed in the California NanoSystems Institute (CNSI). The authors acknowledge useful discussions with Dr. S. Ghosh (UCLA).

**Author Contributions**

A.A.B. led the cooperative project, contributed to the data analysis, and model development. R.K.L. led the model development and analysis. M.T. fabricated the device structures, conducted electronic transport measurements, and contributed to model development and data analysis. N.S. synthesized bulk crystals and performed microscopy and materials characterization. T.S. supervised material synthesis and contributed to materials characterization. J.T. contributed to the characterization of device structures. T.D. contributed to the model development. All authors participated in the manuscript preparation.

**Competing Interests.** The authors declare no competing interests.

**The Data Availability Statement**

The data that support the findings of this study are available from the corresponding author upon reasonable request.

M. Taheri, J. Teeter, T. Debnath, N. Sesing, T. Salguero, R. K. Lake*, and A. A. Balandin* - 2025

**EXTENDED DATA**

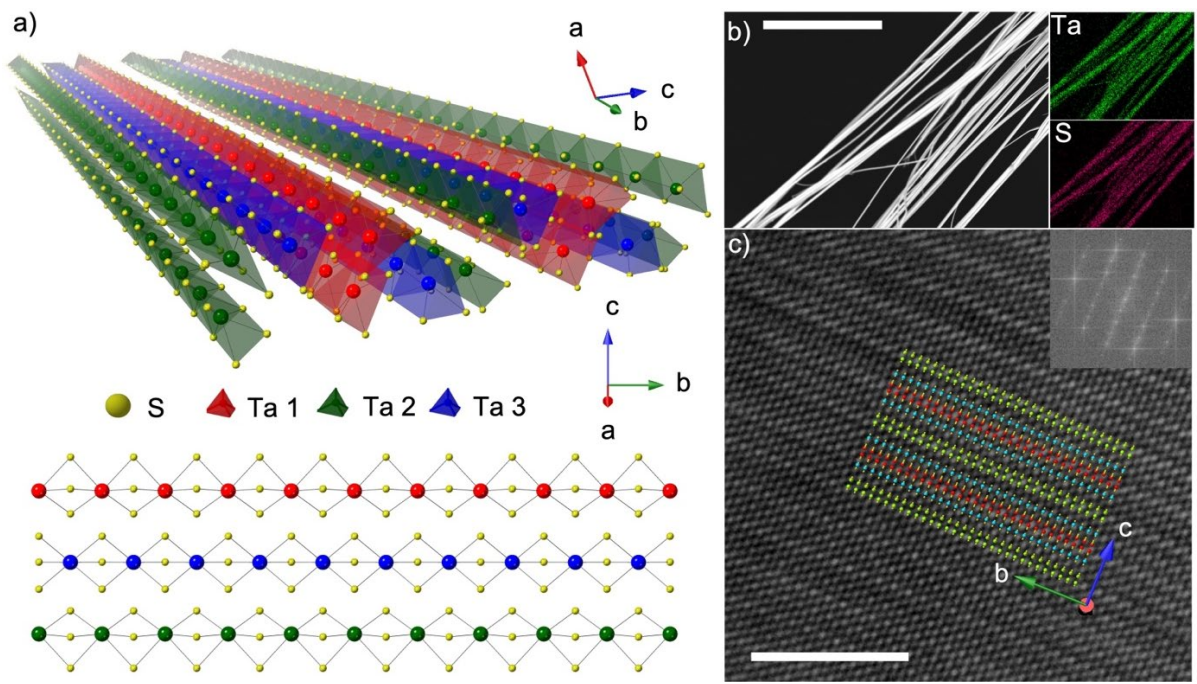

**Extended Data Figure 1: TaS$_3$ crystal structure and material characterization data.** a) An angled perspective view of $o$-TaS$_3$ with C2/m space group (top panel). A view of the three distinct chains, *i.e.*, ball-and-stick model (bottom panel). b) Scanning electron microscopy (SEM) image and energy dispersive x-ray spectroscopy (EDS) maps of tantalum (green) and sulfur (red) showing uniform elemental distribution. The scale bar is 30 $\mu$m. c) High-resolution transmission electron microscopy (HR-TEM) image of TaS$_3$ nanowire with crystal structure overlaid. The scale bar is 3 nm. The inset shows the fast Fourier transform of the HR-TEM image. The characterization data confirms the high quality of the synthesized $o$-TaS$_3$ material.

M. Taheri, J. Teeter, T. Debnath, N. Sesing, T. Salguero, R. K. Lake*, and A. A. Balandin* - 2025

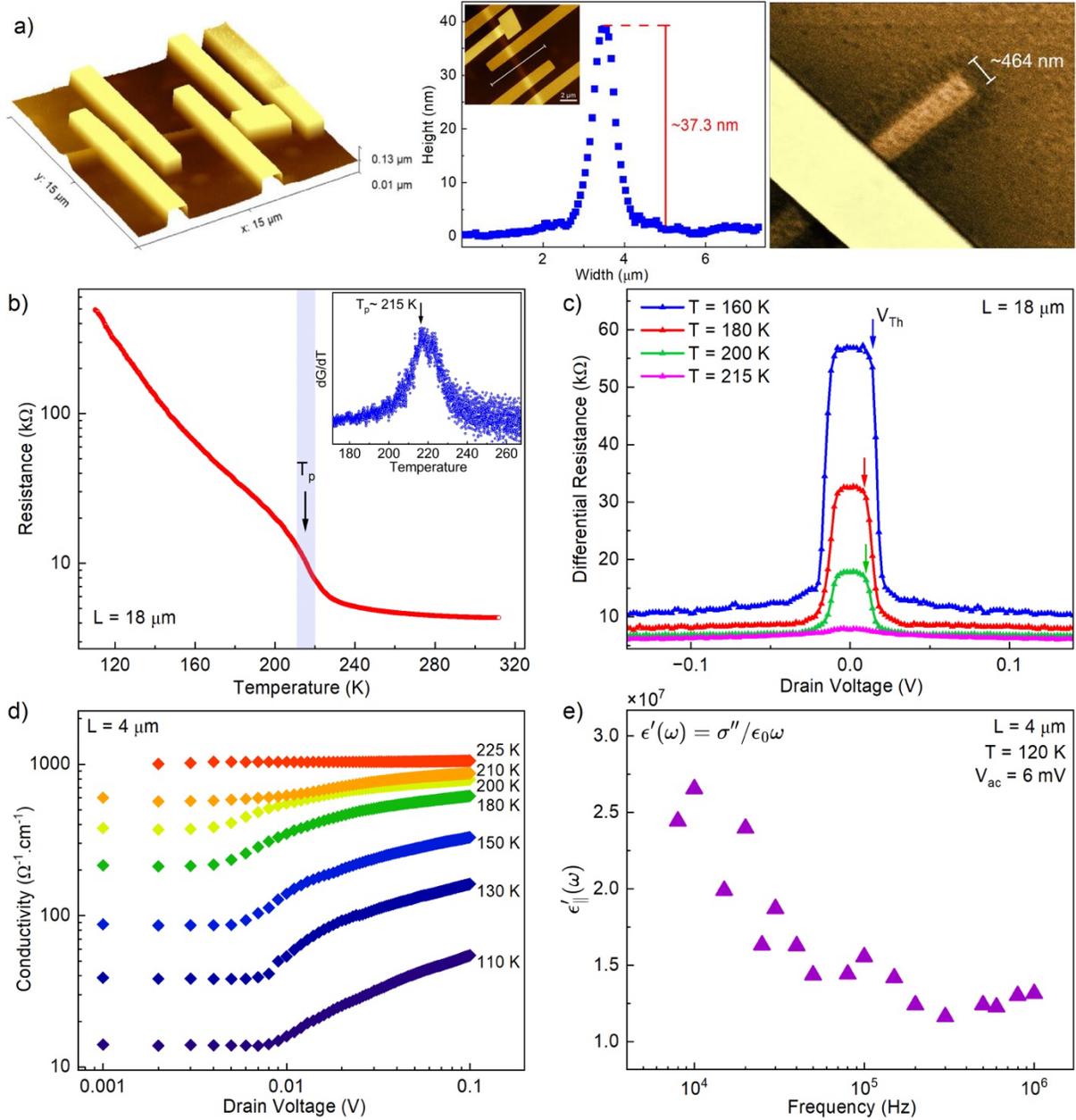

**Extended Data Figure 2: Additional electron transport and characterization data.** a) Atomic force microscopy (AFM) scan of *h*-BN capped *o*-TaS$_3$ device along with the profile view, showing the thickness of the wire to be ~ 37.3 nm. The right panel SEM image shows the width of the channel. b) Temperature dependence of DC resistance with the clear signature of CDW transition at $T_p$ ~ 215 K. c) The differential resistance, $dV/dI$, for a device with the channel length of $L$ = 18 $\mu$m. The arrow shows the threshold voltage. As the temperature approaches $T_p$, the nonlinearity reduces. d) DC conductivity of *o*-TaS$_3$ device with $L$ = 4 $\mu$m as a function of drain bias. The nonlinearity disappears above $T_p$ ~ 215 K (note T = 225 K). The dielectric response of *o*-TaS$_3$ along the chain direction, $\epsilon_\parallel$, at $T$ = 120 K. The giant dielectric constant, ~$10^7$, reflects the polarizability of the CDW condensate in this material.

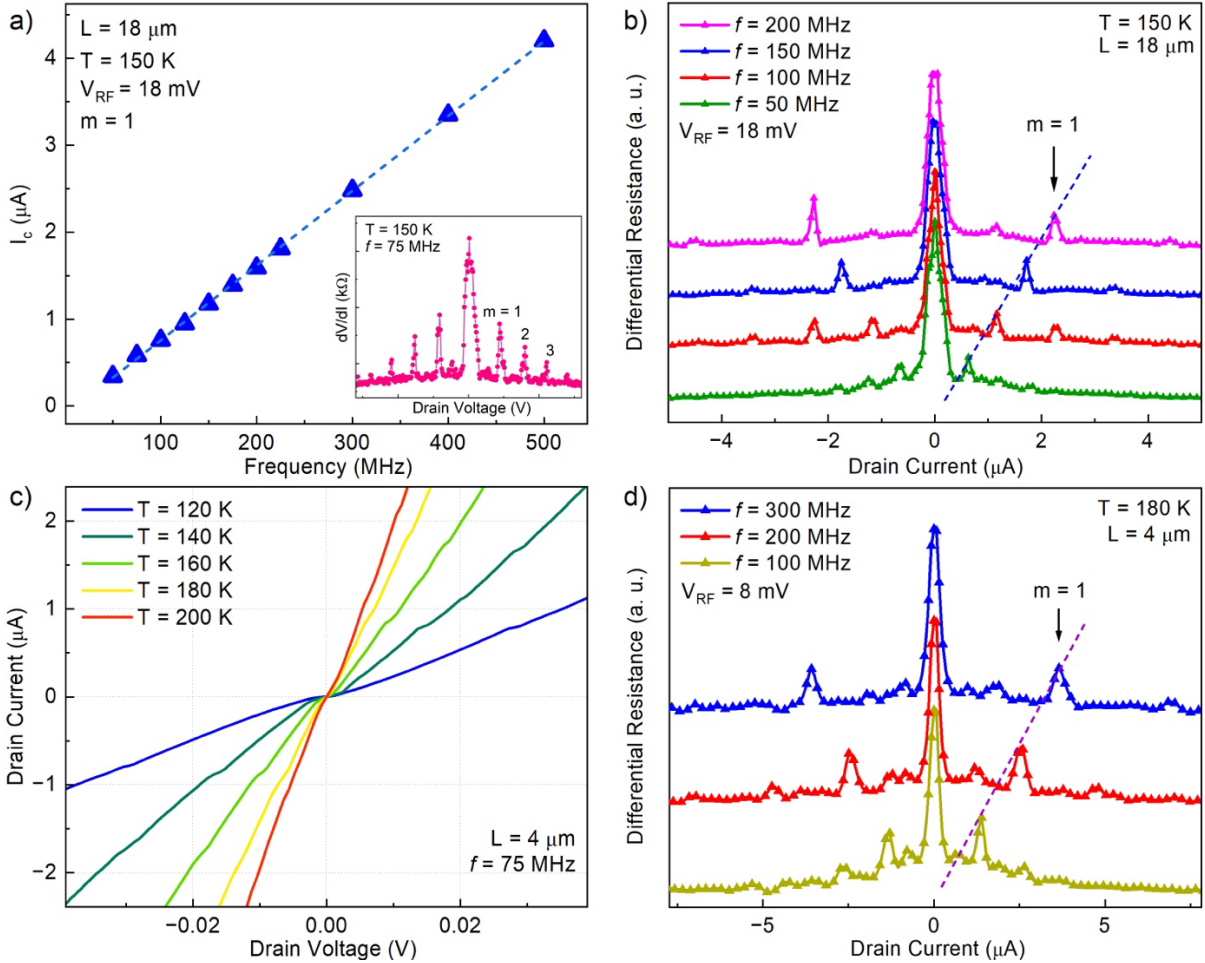

**Extended Data Figure 3: AC-DC coupled measurements of *o*-TaS$_3$ devices.** The CDW current, $I_c$, for a device with $L = 18$ $\mu$m, extracted from the Shapiro peaks (m = 1) at $T = 150$ K. The $I_c$ scale linearly with frequency, which is indicative of a coherent sample. The inset shows the Shapiro peaks for m = 1 and higher harmonics. b) The differential resistance as a function of drain current, revealing the Shapiro peaks with a different frequency drive for $L = 18$ $\mu$m at $T = 150$ K. The dashed line marks the shifts in m = 1 peak as the frequency increases. c) The Shapiro steps, clearly seen in *I-V*s of the device with $L = 4$ $\mu$m under the $f = 75$ MHz radiation for different temperatures $T < T_p$. d) Shapiro peaks in the device with $L = 4$ $\mu$m under AC drive from $f = 100$ MHz to 300 MHz at $T = 180$ K.

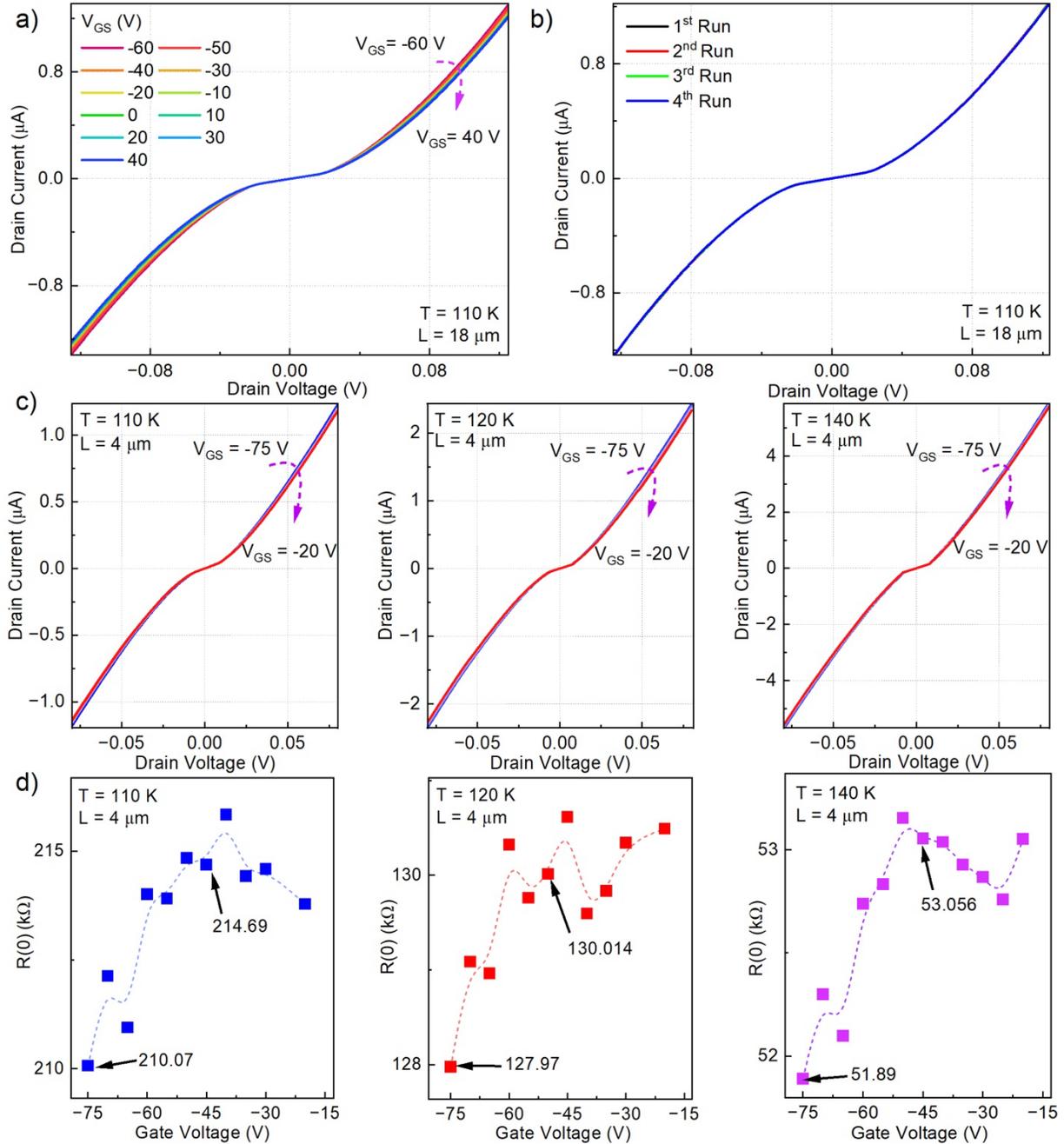

**Extended Data Figure 4: Field-effect in *o*-TaS$_3$ devices.** a) *I-V* characteristics of the device with *L* = 18 μm for the gate voltage range of -60 V to 40 V, at *T* = 110 K. b) *I-V* characteristics after multiple consecutive runs of the same device, without gate bias applied. Note the absence of deviation in current, which further confirms the electrical origin of *I-Vs* and excludes any Joule heating effect for the small source-drain bias voltages. c) Additional output characteristics for the device with *L* = 4 μm measured in the gate bias range of -75 ≤ $V_{GS}$ ≤ -20 and different temperatures. d) The changes in the zero-bias resistance, *R*(0), as a function of $V_{GS}$, which reflect the slight modulation in normal carrier conduction for *L* = 4 μm device.

M. Taheri, J. Teeter, T. Debnath, N. Sesing, T. Salguero, R. K. Lake*, and A. A. Balandin* - 2025

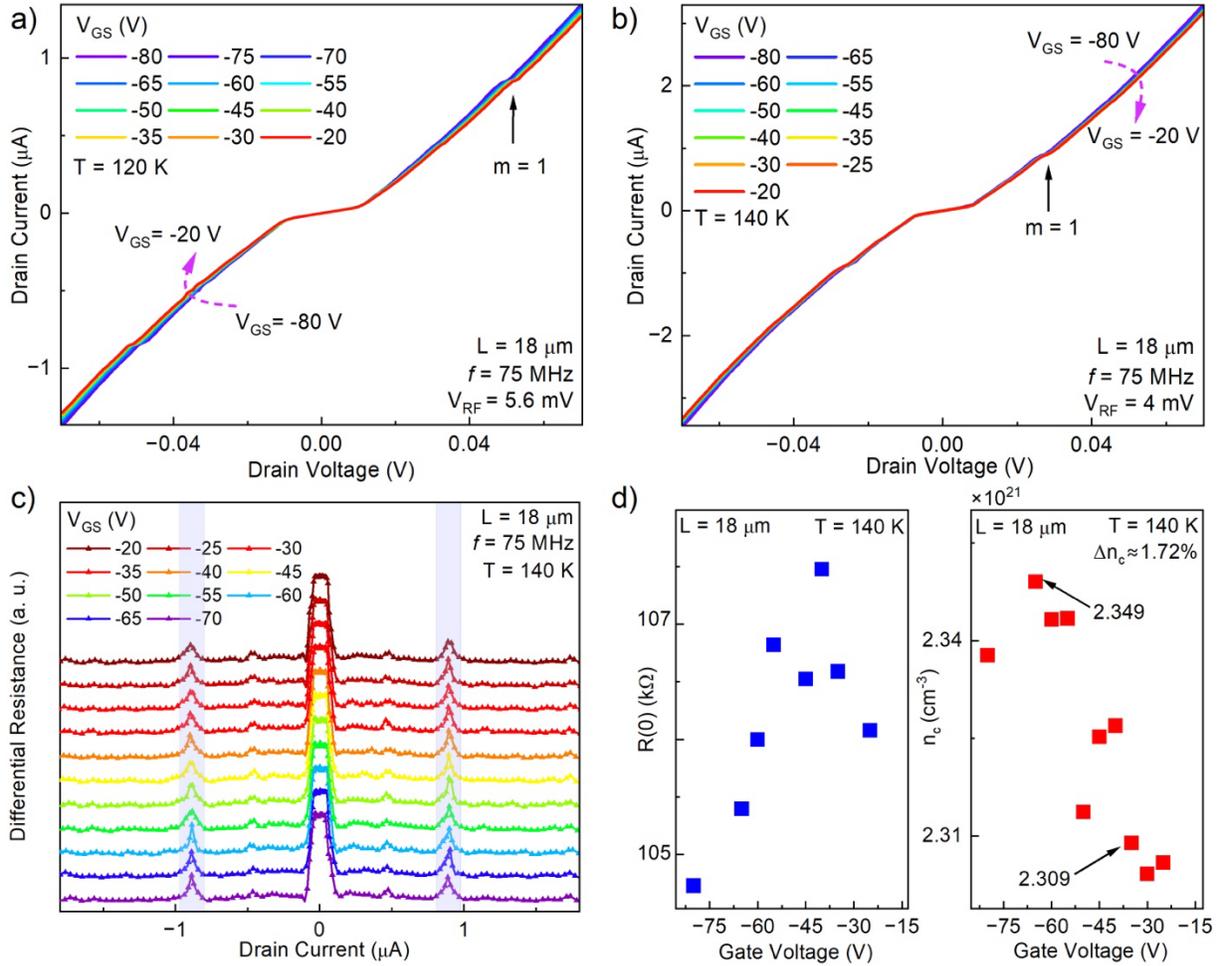

**Extended Data Figure 5: Modulation of CDW charge.** The gate dependence of the Shapiro steps in *I-V*s for the device with $L = 18$ μm at $T = 120$ K (a) and $T = 140$ K (b) under $f = 75$ MHz radiation. c) The *dV/dI* versus drain current shows the changes in m = 1 step (in blue shade) as a function of the gate bias at $T = 140$ K. The left plot in panel (d) represents the slight changes in zero-bias resistance as a function of gate bias for $L = 18$ μm. The gate dependence of $n_c$ for $L = 18$ μm at $T = 140$ K is extracted from Eq. (1) with $\lambda = 4b$ and is shown in right plot of panel (d). The $\Delta n_c \approx 1.72\%$ show the percentage change of $n_c$ between the minimum and maximum of carrier density.